\documentclass[pra,aps,superscriptaddress,groupedaddress,showpacs,twocolumn]{revtex4}
\usepackage{mathrsfs}
\usepackage{xspace,amsmath,amsfonts,amsthm,amssymb,amsbsy}
\usepackage{graphicx}
\usepackage{amssymb}
\usepackage{color}
\usepackage{psfrag}
\usepackage{ifsym}
\usepackage{hyperref}
\usepackage{epstopdf}
\usepackage{units}
\usepackage{bbold}
\usepackage{tikz}
\usepackage{tikz-3dplot}

\definecolor{pink}{rgb}{1,0.078,0.57}

\definecolor{green}{rgb}{0,0.7,0.9}

\newcommand{\ket}[2] {| #1 \rangle_{#2}}
\newcommand{\bra}[2] {\langle #1 |_{#2}}

\newcommand{\bm}{\mathbf{m}}

\newcommand{\br}{\mathbf{r}}

\newcommand{\uv}[1]{\mathbf{\hat{#1}}}

\begin{document}

\title{First-order decomposition of thermal light in terms of 
a statistical mixture of pulses}
\author{Aur\'elia Chenu}
\email[Corresponding author: ]{aurelia.chenu@utoronto.ca}
\affiliation{Department of Chemistry, Centre for
Quantum Information and Quantum Control, 80 Saint George Street, University
of Toronto, Toronto, Ontario, M5S 3H6 Canada}
\author{Agata M. Bra\'nczyk}
\affiliation{Perimeter Institute for Theoretical Physics, Waterloo, Ontario, N2L 2Y5,
Canada}
\author{J. E. Sipe}
\affiliation{Department of Physics, 60 Saint George Street, University of Toronto,
Toronto, Ontario, M5R 3C3 Canada}

\begin{abstract}
We investigate the connection between thermal light and coherent pulses,
constructing mixtures of single pulses that yield the same first-order,
equal-space-point correlation function as thermal light. We present mixtures
involving (i) pulses with a Gaussian lineshape and narrow bandwidths, and
(ii) pulses with a coherence time that matches that of thermal light. We
characterize the properties of the mixtures and pulses. Our results
introduce an alternative description of thermal light in terms of
multi-frequency coherent states, valid for the description of broadband
linear light-matter interactions. We anticipate our results will be relevant
to time-resolved measurements that aim to probe the dynamics of systems as
if they were excited by natural thermal light.
\end{abstract}

\pacs{44.40.+a, 42.50.Ar,  42.50.-p}
\maketitle

\section{Introduction}

The interaction of light with matter triggers fundamental processes in
systems as diverse as bulk semiconductors, quantum dots, photovoltaics, and
living organisms. The study of these processes requires a detailed
understanding of the light used to induce them. Of notable interest is light
from the sun, which plays a major role in the primary production of energy
on the earth through photosynthesis. While the techniques of ultra-fast
spectroscopy are central to studying the dynamics of photo-induced
processes, such as the timescales and mechanisms underlying the initial step
of photosynthesis in light-harvesting complexes, the interpretation of such
experimental results and their relevance to natural excitation conditions
has been questioned, due to differences between laser pulses and thermal
light \cite{Jiang1991a, Mancal2010a, Hoki2011a, Fassioli2012a, Brumer2012a,
Kassal2013a}. 
In principle, systems such as light-harvesting complexes can be completely characterized using ultra-fast 
pulses, and their response to thermal-light excitation can then be inferred theoretically. 
However, understanding the relation of thermal light to pulses of
light will make it possible to directly study the photodynamics of
processes occurring under more natural excitation conditions
 by probing the system with ultra-fast pulses and then appropriately averaging the results according to the light decomposition. 

In Chenu \emph{et al.} \cite{Chenu2014b}, we demonstrated that the state of
thermal light cannot be written as a mixture of single broadband coherent states.
Yet because the experimental characterization of light cannot be done directly but always involves some
interaction with matter, we now focus our attention on states of light that
reproduce some of the \emph{properties} (but not the full density matrix) of
thermal light. We start with the first-order correlation function, which
characterizes effects of light at the level of a linear light-matter
interaction \cite{Glauber1963a}.

A finite-trace mixture of pulses cannot even yield a first-order,
equal-space-point correlation function matching that of thermal light,
except in the unphysical limit that the amplitudes of the pulses would
diverge as the observation volume $\Omega\to \infty$. But
by lifting the requirement of a finite trace and allowing the trace of the
density matrix to scale linearly with the observation volume $\Omega$, we can
mimic this property of thermal light by a mixture of pulses \cite{Chenu2014b}%
. Here we provide the details of how to build these particular mixtures. We
investigate the properties of such mixtures, and characterize the pulses
that compose them. \ After some preliminary considerations in sections IIA
and IIB below, in section IIC we first consider a mixture of pulses with a
Gaussian lineshape, and find that such a decomposition is only possible for
pulses with a FWHM on the order of THz or smaller -- almost \emph{three
orders of magnitude} more narrow than what is physically expected from the
coherence time of thermal radiation at the temperature of the sun. In
section IID we then investigate a mixture without restricting the shape of
the spectral distribution. We find a solution given by a family of pulses
with a lineshape -- and therefore coherence time -- that itself mimics that
of thermal light. \ The mean and spread of the energy and momentum of these
pulses is considered in section IIE. \ Our conclusions, and comments on the
relevance of these studies to problems in linear and nonlinear optics, are
presented in section III.

\section{Matching the first-order correlation function of thermal light}
Recall that the density operator of a thermal state of light at temperature $T$%
, in a normalization volume $V$, is given by $\rho ^{\text{th}}=\mathrm{e}%
^{-\beta H}/\mathrm{Tr}(\mathrm{e}^{-\beta H})$, where $\beta =1/k_{B}T$,
with $k_{B}$ Boltzmann's constant, and $H$ is the field Hamiltonian 
\begin{equation}
H=\sum_{\mathbf{k}\lambda }\,\hbar \,\omega _{k}\,\left(A_{\mathbf{k}\lambda
}^{\dagger }A_{\mathbf{k}\lambda }+\frac{1}{2}\right),
\end{equation}%
here expressed in terms of a sum over modes with wavevector $\mathbf{k}$ and
polarization state $\lambda $; $A_{\mathbf{k}\lambda }^{\dagger }$ and $A_{%
\mathbf{k}\lambda }$ are the creation and annihilation operators, defined
for discrete values of the wave vector, fulfilling the commutation relation $%
[A_{\mathbf{k}\lambda },A_{\mathbf{k}^{\prime }\lambda ^{\prime }}^{\dagger
}]=\delta _{\mathbf{k}\mathbf{k}^{\prime }}\delta _{\lambda \lambda ^{\prime
}}$. In the coherent state basis, the density matrix is given by the
P-representation, 
\begin{equation}
\rho ^{\text{th}}=\prod_{\mathbf{k}\lambda }\int d^{2}\alpha _{\mathbf{k}%
\lambda }C(\alpha _{\mathbf{k}\lambda })|\alpha _{\mathbf{k}\lambda }\rangle
_{{\mathbf{k}\lambda }}\langle \alpha _{\mathbf{k}\lambda }|_{{\mathbf{k}%
\lambda }}\,,  \label{eq:prep}
\end{equation}%
where $C(\alpha _{\mathbf{k}\lambda })=\mathrm{e}^{-|\alpha _{\mathbf{k}%
\lambda }|^{2}/\bar{n}_{\mathbf{k}\lambda }}/(\pi \bar{n}_{\mathbf{k}\lambda
})$, and where $\bar{n}_{\mathbf{k}\lambda }$ is the mean photon number in
mode $\{\mathbf{k},\lambda \}$. Eq. (\ref{eq:prep}) provides a decomposition
of a thermal state in terms of a product of mixtures of \emph{monochromatic}
coherent states $|\alpha _{\mathbf{k}\lambda }\rangle _{{\mathbf{k}\lambda }%
}=\mathrm{e}^{-\nicefrac{|\alpha_{\mathbf{k}\lambda}|^2}{2}}\sum_{n_{\mathbf{%
k}\lambda }}\alpha _{\mathbf{k}\lambda }^{n_{\mathbf{k}\lambda }}/\sqrt{n_{%
\mathbf{k}\lambda }!}\:|n_{\mathbf{k}\lambda }\rangle _{{\mathbf{k}\lambda }}$%
.

Perhaps surprisingly -- or perhaps not, depending on one's perspective -- it
is not possible to construct $\rho ^{th}$ as a mixture of \emph{single} 
\emph{broadband} coherent states, i.e., pulses \cite{Chenu2014b}. However,
by relaxing the constraint that the trace of the density matrix is finite,
we show below how to construct a density matrix from pulses that, while it
is not equal to $\rho ^{\text{th}}$, still leads to a first-order,
equal-space-point correlation function equal to that of thermal light.

\subsection{General definition of a pulse}

To define a pulse of light it is convenient to consider normalization over
all of space. \ That is, we quantize the electromagnetic field using
annihilation and creation operators, ${a}_{\mathbf{k}\lambda }$ and ${a}_{%
\mathbf{k\lambda }}^{\dagger }$ respectively, where the helicity $\lambda $
is positive or negative, the wave vector $\mathbf{k}$ ranges continuously,
and the commutation relations are $\left[ {a}_{\mathbf{k}\lambda },{a}_{%
\mathbf{k}^{\prime }\lambda ^{\prime }}^{\dagger }\right] =\delta (\mathbf{%
k-k}^{\prime })\delta _{\lambda \lambda ^{\prime }}$. We then define a pulse
of light as a classical-like state of the electromagnetic with localized
energy density, characterized by its nominal position $\mathbf{r}_{\mathrm{o}%
}$, a spectral distribution $f_{\mathbf{r}_{\mathrm{o}}s;\mathbf{k\lambda }}$%
, and other parameters \ -- including those necessary to specify the
polarization and the direction of propagation -- that we label collectively
by $s$. The spectral distribution is normalized such that 
\begin{equation}
\sum_{\lambda }\int d\mathbf{k}\;\left\vert f_{\mathbf{r}_{\mathrm{o}}s;%
\mathbf{k}\lambda }\right\vert ^{2}=1,  \label{normalization}
\end{equation}%
with $d\mathbf{k}=dk_{x}dk_{y}dk_{z}$. To build the pulses, we construct
generalized creation operator for a mode defined by the spectral
distribution $f_{\mathbf{r}_{\mathrm{o}}s;\mathbf{k\lambda }}$, 
\begin{equation}
{a}_{\mathbf{r}_{\mathrm{o}}s}^{\dagger }=\sum_{\lambda }\int d\mathbf{k}%
\;f_{\mathbf{r}_{\mathrm{o}}s;\mathbf{k}\lambda }{a}_{\mathbf{k}\lambda
}^{\dagger },
\end{equation}%
with $\left[ {a}_{\mathbf{r}_{\mathrm{o}}s},{a}_{\mathbf{r}_{\mathrm{o}%
}s}^{\dagger }\right] =1$, and describe a pulse by the quantum state 
\begin{equation}  \label{eq:ket_pulse}
\left\vert \alpha _{\mathbf{r}_{\mathrm{o}}s}f_{\mathbf{r}_{\mathrm{o}%
}s}\right\rangle \equiv \mathrm{e}^{\alpha _{\mathbf{r}_{\mathrm{o}}s}{a}_{%
\mathbf{r}_{\mathrm{o}}s}^{\dagger }-\alpha _{\mathbf{r}_{\mathrm{o}%
}s}^{\ast }{a}_{\mathbf{r}_{\mathrm{o}}s}}\left\vert vac\right\rangle ,
\end{equation}%
where $\alpha _{\mathbf{r}_{\mathrm{o}}s}$ is a (complex) amplitude and $%
\left\vert vac\right\rangle $ is the vacuum state, and we use $f_{\mathbf{r}%
_{\mathrm{o}}s}$ to indicate the entire spectral distribution $f_{\mathbf{r}%
_{\mathrm{o}}s;\mathbf{k\lambda }}$. It is easy to confirm that $%
\left\langle \alpha _{\mathbf{r}_{\mathrm{o}}s}f_{\mathbf{r}_{\mathrm{o}%
}s}|\alpha _{\mathbf{r}_{\mathrm{o}}s}f_{\mathbf{r}_{\mathrm{o}%
}s}\right\rangle =1$.

The positive-frequency part of the (Heisenberg) electric field operator can
be written as 
\begin{equation}
{\mathbf{E}}^{(+)}(\mathbf{r},t)=i\sum_{\lambda }\int d\mathbf{k}\;\sqrt{%
\frac{\hbar \omega _{k}}{16\pi ^{3}\epsilon _{0}}}\mathbf{e}_{\mathbf{k}%
\lambda }\mathrm{e}^{i\mathbf{k\cdot r}}\mathrm{e}^{-i\omega _{k}t}{a}_{%
\mathbf{k}\lambda },  \label{fieldpositive}
\end{equation}%
where $\omega _{k}=c\left\vert \mathbf{k}\right\vert $, and we use the
polarization vectors $\mathbf{e}_{\mathbf{k\lambda }}$ for helicity $\lambda 
$, 
\begin{equation} \label{helicity}
\begin{split}
\mathbf{e}_{\mathbf{k}+}& =-\frac{1}{\sqrt{2}}\left( \hat{\mathbf{e}}_{1%
\mathbf{k}}+i\,\hat{\mathbf{e}}_{2\mathbf{k}}\right) , \\
\mathbf{e}_{\mathbf{k}-}& =\frac{1}{\sqrt{2}}\left( \hat{\mathbf{e}}_{1%
\mathbf{k}}-i\,\hat{\mathbf{e}}_{2\mathbf{k}}\right) ,
\end{split}%
\end{equation}%
defined from real orthogonal unit vectors, $\hat{\mathbf{e}}_{1\mathbf{k}}$
and $\hat{\mathbf{e}}_{2\mathbf{k}}$, which fulfill $\hat{\mathbf{e}}_{1%
\mathbf{k}}\times \hat{\mathbf{e}}_{2\mathbf{k}}=\hat{\mathbf{k}}$. The
classical expectation value for the electric field in the state $\left\vert
\alpha _{\mathbf{r}_{\mathrm{o}}s}f_{\mathbf{r}_{\mathrm{o}}s}\right\rangle $%
, 
\begin{equation}
\mathcal{E}^{(\mathbf{r}_{\mathrm{o}}s)}(\mathbf{r},t)\equiv \langle \alpha
_{\mathbf{r}_{\mathrm{o}}s}f_{\mathbf{r}_{\mathrm{o}}s}|_{{}}\mathbf{E}%
^{(+)}(\mathbf{r},t)|\alpha _{\mathbf{r}_{\mathrm{o}}s}f_{\mathbf{r}_{%
\mathrm{o}}s}\rangle _{{}}\,,
\end{equation}%
is then given by (\ref{fieldpositive}) with the operator ${a}_{\mathbf{%
k\lambda }}$ replaced by the complex number $\alpha _{\mathbf{r}_{\mathrm{o}%
}s}f_{\mathbf{r}_{\mathrm{o}}s;\mathbf{k\lambda }}$,%
\begin{equation}
\mathcal{E}^{(\mathbf{r}_{\mathrm{o}}s)}(\mathbf{r},t)=i\sum_{\lambda }\int d%
\mathbf{k}\;\sqrt{\frac{\hbar \omega _{k}}{16\pi ^{3}\epsilon _{0}}}\alpha _{%
\mathbf{r}_{o}s}f_{\mathbf{r}_{o}s;\mathbf{k}\lambda }\mathbf{e}_{\mathbf{k}%
\lambda }\mathrm{e}^{i\mathbf{k\cdot r}}\mathrm{e}^{-i\omega _{k}t}{.}
\label{general pulse}
\end{equation}
\ So the states $\left\vert \alpha _{\mathbf{r}_{\mathrm{o}}s}f_{\mathbf{r}_{%
\mathrm{o}}s}\right\rangle $ characterizing our pulses are multi-chromatic
coherent states, and indeed represents the quantum description of what might
be called \textquotedblleft classical" pulses \cite{Glauber1963a} with a
bandwidth determined by the spectral distribution $f_{\mathbf{r}_{\mathrm{o}%
}s;\mathbf{k\lambda }}$. They are \textquotedblleft coherent" in the sense
that they factorize correlation functions according to ${G}^{(n)(\mathbf{r}_{%
\mathrm{o}}s)}(\mathbf{r}_{1}{t}_{1}\dots \mathbf{r}_{n}{t}_{n};\mathbf{r}_{{%
n}+1}{t}_{{n}+1}\dots \mathbf{r}_{2{n}}{t}_{2{n}})=\prod_{j}\left( \mathcal{E%
}^{(\mathbf{r}_{\mathrm{o}}s)}(\mathbf{r}_{j},{t}_{j})\right) ^{\ast }%
\mathcal{E}^{(\mathbf{r}_{\mathrm{o}}s)}(\mathbf{r}_{j+n},{t}_{j+n})$ for
all orders of ${n}$, as defined by Glauber \cite{Glauber1963a}; here we use
the superscript $(\mathbf{r}_{\mathrm{o}}s)$ on $G$ to identify the state $%
|\alpha _{\mathbf{r}_{\mathrm{o}}s}f_{\mathbf{r}_{\mathrm{o}}s}\rangle _{{}}$%
. \ In particular, for such states we have the first-order correlation
function 
\begin{align}\nonumber
G_{ij}^{(1)(\mathbf{r}_{\mathrm{o}}s)}&(\mathbf{r}_1 t_1;\mathbf{r}_2 t_2){} \\\nonumber
\equiv {}&\bra{
\alpha _{\mathbf{r}_{\mathrm{o}}s}f_{\mathbf{r}_{\mathrm{o}}s}}{} {E}_{i}^{(-)}(\mathbf{r}_1%
,t_1){E}_{j}^{(+)}(\mathbf{r}_2,t_2)\ket{\alpha _{\mathbf{r}_{\mathrm{o}}s}f_{\mathbf{r}%
_{\mathrm{o}}s}}{}  \\
={}&\left( \mathcal{E}_{i}^{(\mathbf{r}_{\mathrm{o}}s)}(\mathbf{r}_1,{t}_1)\right) ^{\ast }%
\left( \mathcal{E}_{j}^{(\mathbf{r}_{\mathrm{o}}s)}(\mathbf{r}_2,{t}_2) \right),\label{eq:G1_pulse}
\end{align}
where subscripts on field labels indicate Cartesian components,
i.e. $ \mathcal{E}_{i}^{(\mathbf{r}_{\mathrm{o}}s)}(\mathbf{r},{t}%
) =  \mathcal{E}^{(\mathbf{r}_{\mathrm{o}}s)}(\mathbf{r},{t}%
) \cdot  \uv{i}$. \ 

We consider families of pulses such that, for a fixed set of properties $s$,
the pulses only differ by their nominal position $\mathbf{r}_{\mathrm{o}}$.
For such a family of pulses we have 
\begin{equation}
f_{\mathbf{r}_{\mathrm{o}}s;\mathbf{k\lambda }}=K(s,\mathbf{k}\lambda )\,%
\mathrm{e}^{-i\mathbf{k\cdot r}_{\mathrm{o}}},  \label{eq:f_general}
\end{equation}%
and the associated $\mathcal{E}^{(\mathbf{r}_{\mathrm{o}}s)}(\mathbf{r},{t})$
depends on $\mathbf{r}$ and $\mathbf{r}_{\mathrm{o}}$ only through its
dependence on $(\mathbf{r-r}_{\mathrm{o}})$; we will give particular
examples of $K(s,\mathbf{k\lambda )}$ below, but for the moment we keep the
function very general. \ Nonetheless, we do assume that each member of the
family is well localized in space at some nominal time $t=0$, with $%
G_{ij}^{(1)(\mathbf{r}_{\mathrm{o}}s)}(\mathbf{r}0;\mathbf{r}0)\rightarrow 0$
as $\left\vert \mathbf{r-r}_{\mathrm{o}}\right\vert \rightarrow \infty $,
and that the integral of $G_{ij}^{(1)(\mathbf{r}_{\mathrm{o}}s)}(\mathbf{r}0;%
\mathbf{r}0)$ over all space is finite. \ Then for fixed $\mathbf{r}$ the
integral over all $\mathbf{r}_{\mathrm{o}}$ of $G_{ij}^{(1)(\mathbf{r}_{%
\mathrm{o}}s)}(\mathbf{r}0;\mathbf{r}0)$ will also be finite. \ 

\subsection{Construction of a statistical mixture of pulses}

In Chenu \emph{et al.} \cite{Chenu2014b}, we suggested the consideration 
of \emph{trace-improper} density operators, i.e., with a trace diverging as the observation volume $\Omega \to \infty$.
Here we explicitly consider the observation volume spanning all of space, and write $\rho^{\rm imp}$ as
\begin{equation}   \label{rhoimp}
\rho ^{\mathrm{imp}}=\int ds\int d\mathbf{r}_{\mathrm{o}}\;\bar{p}%
(s)\left\vert \alpha _{\mathbf{r}_{\mathrm{o}}s}f_{\mathbf{r}_{\mathrm{o}%
}s}\right\rangle \left\langle \alpha _{\mathbf{r}_{\mathrm{o}}s}f_{\mathbf{r}%
_{\mathrm{o}}s}\right\vert
\end{equation}%
where the integral over $\br_{\rm o}$ ranges over all of space, 
where $\bar{p}(s)\geq 0$, and 
\begin{equation}
\int ds\;\bar{p}(s)=\frac{1}{\mathcal{V}}.
\end{equation}%
Here $\mathcal{V}$ is a constant with units of volume. The non-negative
distribution $\bar{p}(s)$ has dimension of $[\Omega^{-1}V_{s}^{-1}]$, where $%
V_{s} $ is the volume of the integration space of the parameters $s$. \
The issue we wish to
address here is: \ How do we build such a mixture of pulses, characterized
by $\{\bar{p}(s),f_{\mathbf{r}_{\mathrm{o}}s}\}$, that correctly describes
the first order, equal-space-point correlation function of thermal light, $%
G_{ij}^{(1)\mathrm{th}}(\mathbf{r}t;\mathbf{r}{t^{\prime }})$? \ 
That correlation function is given (\cite%
{Loudon2000},\cite{Kano1962a}) by 
\begin{eqnarray}
G_{ij}^{(1)\mathrm{th}}(\mathbf{r}t;\mathbf{r}{t^{\prime }}) &=&\mathrm{Tr}%
\left( \rho ^{\mathrm{th}}{E}_{i}^{(-)}(\mathbf{r},t){E}_{j}^{(+)}(\mathbf{r}%
,{t^{\prime }})\right)  \label{thermal1} \\
&=&\delta _{ij}\int_{0}^{\infty }\frac{\hbar ck^{3}}{6\pi ^{2}\epsilon _{0}}%
\frac{\mathrm{e}^{-ick(t^{\prime }-t)}}{\mathrm{e}^{\beta \hbar ck}-1}dk. 
\notag
\end{eqnarray}

To proceed, we take the set of parameters $s$ used to define our pulses to
include a central wave vector $\mathbf{k}_{\mathrm{o}}=k_{\mathrm{o}}\hat{\mathbf{m}}$ of the pulse, 
with the unit vector $\uv{m}$ identifying the main propagation direction, 
and a
unit vector $\hat{\mathbf{n}}$ characterizing the polarization; thus $s=\{%
k_{\mathrm{o}},\uv{m},\uv{n}\}$. \ Next we specify a form for $%
f_{\mathbf{r}_{\mathrm{\mathrm{o}}}s;\mathbf{k\lambda }}$ (\ref{eq:f_general}%
), and describe our pulses by 
\begin{equation}
f_{\mathbf{r}_{\mathrm{o}}s;\mathbf{k\lambda }}=\mathcal{N\;}L(\mathbf{k,k}_{%
\mathrm{\mathrm{o}}})\,(\mathbf{e}_{\mathbf{k\lambda }}^{\ast })\cdot (%
\mathbf{k\times \hat{n}})\,\mathrm{e}^{-i\mathbf{k\cdot r}_{\mathrm{o}}},
\label{eq:K_general}
\end{equation}%
where $L(\mathbf{k,k}_{\mathrm{\mathrm{o}}})$ is a real function, and $%
\mathcal{N}$ is a normalization constant chosen such that $f_{\mathbf{r}_{%
\mathrm{\mathrm{o}}}s;\mathbf{k\lambda }}$ is normalized. \ The convenience
of this form can be most easily seen by looking at the expectation value (%
\ref{general pulse}) of the positive-frequency part of the electric field
operator $\mathbf{E}^{(+)}(\mathbf{r},t)$ for such a pulse, 
\begin{eqnarray}\label{eq:E_general}
\mathcal{E}^{(\mathbf{r}_{\mathrm{\mathrm{o}}}s)}(\mathbf{r},t)&=& i\mathcal{N%
}\alpha _{\mathbf{r}_{\mathrm{\mathrm{o}}}s}\int d\mathbf{k}\;\sqrt{\frac{%
\hbar \omega _{k}}{16\pi ^{3}\epsilon _{0}}}(\mathbf{k\times \hat{n}})L(%
\mathbf{k,k}_{\mathrm{\mathrm{o}}}) \nonumber \\ 
&\:& \times  \mathrm{e}^{i\mathbf{k\cdot (r-r}_{\mathrm{\mathrm{o}}})}\mathrm{e}%
^{-i\omega _{k}t}\,,
\end{eqnarray}%
where we have used 
\begin{equation} \label{kcrossn}
\sum_{\lambda }(\mathbf{k\times \hat{n})\cdot e}_{\mathbf{k\lambda }}^{\ast }%
\mathbf{e}_{\mathbf{k\lambda }}=(\mathbf{k\times \hat{n}),}
\end{equation}%
which holds since $(\mathbf{k\times \hat{n})}$ is necessarily perpendicular
to $\mathbf{k}$ and 
\begin{equation}\label{polsum}
\sum_{\lambda }\mathbf{e}_{\mathbf{k\lambda }}^{\ast }\mathbf{e}_{\mathbf{%
k\lambda }}=\mathbf{U-\hat{k}\hat{k},}
\end{equation}%
where $\mathbf{U}$ is the unit dyadic. \ Note that while the use of any
spectral function $f_{\mathbf{r}_{\mathrm{o}}s;\mathbf{k\lambda }}$ in (\ref%
{general pulse}) would lead to a divergenceless expectation value $\mathcal{E%
}^{(\mathbf{r}_{\mathrm{\mathrm{o}}}s)}(\mathbf{r},t)$, the form (\ref%
{eq:K_general}) leads to an expression (\ref{eq:E_general}) for $\mathcal{E}^{(%
\mathbf{r}_{\mathrm{\mathrm{o}}}s)}(\mathbf{r},t)$ in which the polarization
vectors $\mathbf{e}_{\mathbf{k\lambda }}$ do not appear. \ Further, we see
that $\mathcal{E}^{(\mathbf{r}_{\mathrm{\mathrm{o}}}s)}(\mathbf{r},t)$ is
centered in space at $\mathbf{r}_{\mathrm{o}}$, and the direction of the
vector field $\mathcal{E}^{(\mathbf{r}_{\mathrm{\mathrm{o}}}s)}(\mathbf{r}%
,t) $ lies exclusively in the plane perpendicular to $\mathbf{\hat{n}}$ such
that $\hat{\mathbf{n}}\cdot \mathcal{E}^{(\mathbf{r}_{\mathrm{\mathrm{o}}%
}s)}(\mathbf{r},t)=0$. \ Thus $\mathbf{\hat{n}}$ identifies the polarization
of the field $\mathcal{E}^{(\mathbf{r}_{\mathrm{\mathrm{o}}}s)}(\mathbf{r}%
,t) $, in that it defines the direction in which the pulse expectation value
has \textit{no }component. \ 

The correlation function for each pulse is given by (\ref{eq:G1_pulse}).
\ From the definition of $\rho ^{\mathrm{imp}}$ in (\ref{rhoimp}), we
see that the correlation function for the mixture of pulses is given by the
weighted sum of correlation functions for individual pulses. Considering the
choice of parameters $s$, we explicitly write the integral over the set of
parameters as $\int ds\rightarrow \int dk_{\mathrm{o}} d\uv{m} d\hat{\mathbf{%
n}}$. In addition, we assume a uniform distribution over pulse directions $\uv{m}$ 
and,\ for a given $\uv{m}$, we consider an equal weighting of all $\mathbf{\hat{n}}$ perpendicular
to $\uv{m}$. Thus the probability distribution $\bar{p%
}(s)$ depends only on the magnitude of the central wave vector, $\bar{p}%
(s)\rightarrow p(k_{\mathrm{o}})$, and the correlation function for the
mixture becomes 
\begin{equation}
G_{ij}^{(1)\mathrm{imp}}(\mathbf{r}t;\mathbf{r}t^{\prime })={}\int dk_{\mathrm{o}}%
d\uv{m}d\hat{\mathbf{n}}\int d\mathbf{r}_{\mathrm{o}}\:{p}(k_{%
\mathrm{o}})\,G_{ij}^{(1)(\mathbf{r}_{\mathrm{o}}s)}(\mathbf{r}t;\mathbf{r}%
t^{\prime })\,.  \label{eq:G1_mixture-general}
\end{equation}

In the rest of this section, we investigate two different mixtures for which 
$G_{ij}^{(1)\mathrm{imp}}(\mathbf{r}t;\mathbf{r}t^{\prime })=G_{ij}^{(1)%
\mathrm{th}}(\mathbf{r}t;\mathbf{r}t^{\prime })$, and provide
characteristics of the pulses involved. \ We first start with a mixture
constrained by adopting a Gaussian form for $L(\mathbf{k,k}_{\mathrm{o}})$,
and define the conditions under which the $G_{ij}^{(1)\mathrm{imp}}(\mathbf{r%
}t;\mathbf{r}t^{\prime })$ that results can represent $G_{ij}^{(1)\rm th}(%
\mathbf{r}t;\mathbf{r}t^{\prime })$. Considering the restrictions found on
the bandwidth of pulses that successfully fulfil this decomposition, we then
lift the restriction of $L(\mathbf{k,k}_{\mathrm{o}})$ to a Gaussian and
propose another form.

\subsection{Mixture of Gaussian-like pulses}

We take the pulses to be characterized by a Gaussian form for $L(\mathbf{k,k}%
_{\mathrm{o}})$, and label all properties related to this particular
lineshape with the superscript `$g$'. The spectral distribution (\ref%
{eq:K_general}) is given by 
\begin{equation}
f_{\mathbf{r}_{\mathrm{o}}s;\mathbf{k\lambda }}^{g}=\mathcal{N}^{g}\mathcal{%
\;}L^{g}(\mathbf{k,k}_{\mathrm{\mathrm{o}}})\,(\mathbf{e}_{\mathbf{k\lambda }%
}^{\ast })\cdot (\mathbf{k\times \hat{n}})\,\mathrm{e}^{-i\mathbf{k\cdot r}_{%
\mathrm{o}}},  \label{spectralGaussian}
\end{equation}%
where we define the real function 
\begin{equation}
L^{g}(\mathbf{k},\mathbf{k}_{\mathrm{o}})=\mathrm{e}^{-\frac{|\mathbf{k}-%
\mathbf{k}_{\mathrm{o}}|^{2}}{2\sigma ^{2}}}.  \label{eq:f_Gaussian}
\end{equation}%
The normalization condition (\ref{normalization}) on $f_{\mathbf{r}_{\mathrm{%
o}}s;\mathbf{k\lambda }}^{g}$ (\ref{eq:f_general}) then leads to the
normalization constant: 
\begin{equation}
\mathcal{N}^{g}=\left( {\pi \sqrt{\pi }\sigma ^{3}(k_{\mathrm{o}}^{2}+\sigma
^{2})}\right) ^{-\frac{1}{2}},  \label{eq:n_Gaussian}
\end{equation}%
and the expectation value of the positive-frequency part of the electric
field in such a pulse (\ref{eq:E_general}) is given by 
\begin{eqnarray}\label{E_gaussian}
\mathcal{E}^{g}(\mathbf{r},t)&=& i\mathcal{N}^{g}\alpha _{\mathbf{r}_{\mathrm{%
\mathrm{o}}}s}\int d\mathbf{k}\;\sqrt{\frac{\hbar \omega _{k}}{16\pi
^{3}\epsilon _{0}}}(\mathbf{k\times \hat{n}})L^{g}(\mathbf{k,k}_{\mathrm{%
\mathrm{o}}}) \nonumber \\
&\:&\times \mathrm{e}^{i\mathbf{k\cdot (r-r}_{\mathrm{\mathrm{o}}})}\mathrm{e}%
^{-i\omega _{k}t}\,,
\end{eqnarray}%
where the dependence of $\mathcal{E}^{g}(\mathbf{r},t)$ on $\mathbf{r}_{%
\mathrm{o}}$ and $s$ is kept implicit.

The first-order correlation function for the mixture (\ref%
{eq:G1_mixture-general}) of such pulses is 
\begin{equation}
\begin{split}
G_{ij}^{(1){g}}(\mathbf{r}t;\mathbf{r}t^{\prime })={}& \int dk_{%
\mathrm{o}} d\uv{m}d\hat{\mathbf{n}}\int d\mathbf{r}_{\mathrm{o}}\:p(k_{\mathrm{o}%
}) \\
& \times \left( \mathcal{E}_{i}^{g}(\mathbf{r},t)\right) ^{\ast }\,\left( 
\mathcal{E}_{j}^{g}(\mathbf{r},t^{\prime })\right) \,.
\end{split}
\label{gaussianG}
\end{equation}%
When expressions (\ref{E_gaussian}) for $\mathcal{E}_{i}^{g}(\mathbf{r},t)$ and $%
\mathcal{E}_{j}^{g}(\mathbf{r},t)$ are substituted into (\ref{gaussianG})
there will be integrals over two wavevector variables, but the integral over 
$\mathbf{r}_{\mathrm{o}}$ will
reduce it to an integral over only one remaining such variable $\mathbf{k}$.
\ 
We do the integral over $\mathbf{\hat{n}}$
for a fixed $\mathbf{\hat{m}}$, and then do the integrals over the direction 
$\mathbf{\hat{m}}$ and the direction of $\mathbf{k}$. \ Details are given in
Appendix A; the result is that (\ref{gaussianG}) can be written in terms of
only integrals over the magnitudes of $\mathbf{k}_{\mathrm{o}}$ and $\mathbf{%
k}$, $k_{\mathrm{o}}$ and $k$ respectively: \ 
\begin{equation}
\begin{split}
G_{ij}^{(1){g}}(\mathbf{r}t;\mathbf{r}(t+\tau ))={}& \delta _{ij}|\alpha
|^{2}\int_{0}^{\infty }d{k}\int_{0}^{\infty }d{k}_{\mathrm{\mathrm{o}}}\:p({k}%
_{\mathrm{\mathrm{o}}}) \\
\times & \frac{\hbar \,c\,k^{3}}{6\pi ^{2}\epsilon _{0}}M({k},{k}_{\mathrm{%
\mathrm{o}}})\,\mathrm{e}^{-i{c}{k}\tau }\,,
\end{split}
\label{Gaussianresult}
\end{equation}%
with 
\begin{equation}
\begin{split}
M(k,k_{\mathrm{o}})\equiv & \frac{4\pi ^{3}\sqrt{\pi }\sigma }{a(\sigma
^{2}+k_{\mathrm{o}}^{2})}\,\left( (a^{2}-a+1)\,\mathrm{e}^{-\left( \frac{%
k-k_{\mathrm{o}}}{\sigma }\right) ^{2}}\right. \\
-& \left. (a^{2}+a+1)\,\mathrm{e}^{-\left( \frac{k+k_{\mathrm{o}}}{\sigma }%
\right) ^{2}}\right) \,,
\end{split}%
\end{equation}%
where $a=2kk_{\mathrm{o}}/\sigma ^{2}$.

We can build such a mixture with a correlation function $G_{ij}^{(1)g}%
(\mathbf{r}t;\mathbf{r}(t+\tau ))$ equal to that of thermal light if there
exists a non-negative function $p(k_{\mathrm{o}})$ fulfilling 
\begin{equation}
\int_{0}^{\infty }p({k}_{\mathrm{\mathrm{o}}})\,M({k},{k}_{\mathrm{\mathrm{o}%
}})\,d{k}_{\mathrm{\mathrm{o}}}={}\frac{\bar{n}_{\mathbf{k}\lambda }}{%
|\alpha |^{2}}\,.  \label{eq:dgsfgd}
\end{equation}%
It is possible to search for solutions of (\ref{eq:dgsfgd}) numerically,
and we find a solution for pulses with a spectral FWHM on the order of THz
or smaller, i.e., pulses that are on the order of picoseconds in length or
longer. But no physical solutions, with positive definite distributions $%
p(k_{\mathrm{o}})$, can be found for pulses with a bandwidth as broad as the
thermal spectrum, i.e. femtosecond pulses.

These pulses have a surprisingly narrow bandwidth, almost \emph{three orders
of magnitude} narrower than what is physically expected from the coherence
time of the thermal radiation. The problem is that the Gaussian shape (\ref%
{eq:f_Gaussian}) differs too much from the shape required to guarantee that
the norm of the integrand of (\ref{thermal1}) is reproduced. Thus the only
way to reproduce the thermal $G^{(1) \rm th}(\mathbf{r}t;\mathbf{r}(t+\tau ))$ is
to choose the width $\sigma $ in Fourier space so small that, compared with
the thermal spectrum, $L^{g}(\mathbf{k},\mathbf{k}_{\mathrm{o}})$ is
essentially proportional to a Dirac delta function; then the function $p(k_{%
\mathrm{o}})$ itself is relied on to capture the shape of that integrand. \
Minor modifications of (\ref{eq:f_Gaussian}) could be considered, such as
multiplying or dividing $L^{g}(\mathbf{k,k}_{\mathrm{o}})$ by powers of $k$;
but we would expect this conclusion to hold even with such changes.

In an attempt to find a mixture consisting of pulses with a broader
bandwidth, we lift the restriction on the shape of the spectral distribution.
We find pulses that themselves match the spectrum of thermal light.

\subsection{Mixture of pulses with unrestricted line shape: `thermal' pulses}

We now investigate a mixture of pulses without initially restricting the
shape of the spectral distribution (\ref{eq:f_general}). We consider all
pulses to have the same lineshape and suppose that the set of parameters $s$
depends on the nominal wave vector $\mathbf{k}_{\mathrm{o}}$ only by
depending on its direction, represented again by the unit vector $\hat{%
\mathbf{m}}$. Each pulse is then characterized by the set of parameters $s=\{%
\hat{\mathbf{m}},\hat{\mathbf{n}}\}$; consequently $V_{s}$ is dimensionless
and $\bar{p}(s)$ has dimension of $\Omega^{-1}$. \ In our mixture we will
integrate over all possible $\mathbf{\hat{m}}$ and all allowable $\mathbf{%
\hat{n}}$, and so $\bar{p}(s)\rightarrow p$, a constant.

We label these pulses with the superscript $b$ for `broadband' and specify
the spectral distribution (\ref{eq:K_general}) by%
\begin{equation}
f_{\mathbf{r}_{\mathrm{o}}s;\mathbf{k\lambda }}^{b}=\mathcal{N}^{b}\mathcal{%
\;}L^{b}(\mathbf{k,k}_{\mathrm{\mathrm{o}}})\,(\mathbf{e}_{\mathbf{k\lambda }%
}^{\ast })\cdot (\mathbf{k\times \hat{n}})\,\mathrm{e}^{-i\mathbf{k\cdot r}_{%
\mathrm{o}}}  \label{spectralbroadband}
\end{equation}%
with%
\begin{equation}
L^{b}(\mathbf{k},\mathbf{k}_{\mathrm{o}})=l(k)\,\upsilon (\hat{\mathbf{k}}%
\cdot \hat{\mathbf{m}}),  \label{f_broadband}
\end{equation}%
where the function $\upsilon (x)$ is chosen to characterize the spread in
the direction of wave vectors in the pulse and should be peaked at $x=1$ for 
$\hat{\mathbf{m}}$ to represent the nominal direction of propagation of the
pulse; the function $l(k)$ is now relied on to help capture the shape of the
norm of the integrand in (\ref{thermal1}). \ The normalization condition
(\ref{normalization}) on $f_{\mathbf{r}_{\mathrm{o}}s;\mathbf{k}\lambda }$
leads to the normalization constant 
\begin{equation}
\mathcal{N}^{b}=\left[ \pi (C_{0}+C_{2})\int_{0}^{\infty }k^{4}l(k)^{2}dk%
\right] ^{-\frac{1}{2}},  \label{n_broadband}
\end{equation}%
where 
\begin{equation}
C_{n}=\int_{-1}^{1}dx\,\,x^{n}\upsilon ^{2}(x),  \label{Cndef}
\end{equation}%
and the expectation value of the positive-frequency part of the electric
field in such a pulse (\ref{eq:E_general}) is given by 
\begin{equation}
\begin{split}
\mathcal{E}^{b}(\mathbf{r},t)=& i\mathcal{N}^{b}\alpha _{\mathbf{r}_{\mathrm{%
\mathrm{o}}}s}\int d\mathbf{k}\;\sqrt{\frac{\hbar \omega _{k}}{16\pi
^{3}\epsilon _{0}}}(\mathbf{k\times \hat{n}})L^{b}(\mathbf{k,k}_{\mathrm{%
\mathrm{o}}}) \\
&\times \mathrm{e}^{i\mathbf{k\cdot (r-r}_{\mathrm{\mathrm{o}}})}\mathrm{e}%
^{-i\omega _{k}t}\,,
\end{split}
\label{E_broadband}
\end{equation}%
where the dependence of $\mathcal{E}^{b}(\mathbf{r},t)$ on $\mathbf{r}_{%
\mathrm{o}}$ and $s$ is kept implicit. \ 

The first-order, equal-space-point correlation function for such a mixture
of pulses is 
\begin{equation}
G_{ij}^{(1){b}}(\mathbf{r}t;\mathbf{r}t^{\prime })={}\int d\hat{\mathbf{m}}d%
\hat{\mathbf{n}}\int d\mathbf{r}_{\mathrm{o}}\:p\left( \mathcal{E}_{i}^{b}(%
\mathbf{r},t)\right) ^{\ast }\,\left( \mathcal{E}_{j}^{b}(\mathbf{r}%
,t^{\prime })\right) \,.  \label{eq:g1_b-def}
\end{equation}%
As in the mixture of Gaussian pulses, when expressions for $\mathcal{E}%
_{i}^{b}(\mathbf{r},t)$ and $\mathcal{E}_{j}^{b}(\mathbf{r},t)$ are
substituted into (\ref{eq:g1_b-def}) there will be integrals over two
wavevector variables, but the integral
over $\mathbf{r}_{\mathrm{o}}$ will reduce it to an integral over only one
remaining such variable $\mathbf{k}$. \ The integrals over $\mathbf{\hat{n}}$%
, $\mathbf{\hat{m}}$, and $\mathbf{\hat{k}}$ can then be done, as detailed
in Appendix B, and we find 
\begin{equation}
\begin{split}
G_{ij}^{(1){b}}(\mathbf{r}t;\mathbf{r}(t+\tau ))=& \delta _{ij}\frac{%
p|\alpha |^{2}}{\int_{0}^{\infty }k^{4}l(k)^{2}dk}\frac{4\pi ^{2}}{3\epsilon
_{0}} \\
&\times  \int_{0}^{\infty }\hbar ck^{5}l(k)^{2}\mathrm{e}^{-ick\tau }dk\,.
\end{split}
\label{broadbandresult}
\end{equation}%
\ 

Imposing the condition $G_{ij}^{(1){b}}(\mathbf{r}t;\mathbf{r}(t+\tau
))=G_{ij}^{(1)\mathrm{th}}(\mathbf{r}t;\mathbf{r}(t+\tau ))$ yields the
condition%
\begin{equation}
l(k)=k^{-1}(\mathrm{e}^{\beta \hbar ck}-1)^{-\frac{1}{2}},  \label{luse}
\end{equation}%
with then 
\begin{equation}
p|\alpha |^{2}=\frac{1}{8\pi ^{4}}\int_{0}^{\infty }\frac{k^{2}}{\mathrm{e}%
^{\beta \hbar ck}-1}dk=\frac{4\zeta (3)}{\pi ^{4}(\beta c\hbar )^{3}}.
\label{eq:p_th_pulse}
\end{equation}%
This mixture consists of pulses that only differ in their position in space,
their propagation direction, and their polarization; they are weighted
identically. The coherence time of the pulse, as determined by its spectral
FWHM, is the same as thermal light, which is approximatively $1.3$ fs for $%
T=5777$ $K$. The pulse energy is restricted by the condition that the
product $p|\alpha |^{2}$, where recall $|\alpha |^{2}$ corresponds to the
average number of photons in a coherent state, is fixed.

Because of the link of their spectrum to the thermal spectrum, we refer to
these pulses as ``thermal pulses".

\subsection{Properties of the thermal pulses}

In this section we consider some elementary properties of these thermal
pulses, represented by kets (\ref{eq:ket_pulse}) $\left\vert \alpha _{%
\mathbf{r}_{\mathrm{o}}s}f_{\mathbf{r}_{\mathrm{o}}s}^{b}\right\rangle $, in
the special case where the elements (\ref{eq:K_general}) of $f_{\mathbf{r}_{%
\mathrm{o}}s}^{b}$ are specified by $L^{b}$ (\ref{f_broadband}) and $%
\mathcal{N}^{b}$ (\ref{n_broadband}). \ Since these are the only pulses we
consider in this section we avoid clutter in the notation by writing these
kets simply as $\left\vert \alpha f\right\rangle $. \ 

The operator characterizing the energy above the vacuum is 
\begin{equation}
\mathfrak{E}=\sum_{\lambda }\int d\mathbf{k\;}\hbar \omega _{k}n_{\mathbf{k}%
\lambda },
\end{equation}%
where $n_{\mathbf{k\lambda }}=a_{\mathbf{k\lambda }}^{\dagger }a_{\mathbf{k}%
\lambda }$, and so the expectation value of the energy of one of our pulses
is easily found to be, 
\begin{equation}
\begin{split}
\left\langle \mathfrak{E}\right\rangle =& \sum_{\lambda }\int d\mathbf{k}%
\langle \alpha f|\hbar \omega _{k}{n}_{\mathbf{k}\lambda }|\alpha f\rangle
_{{}} \\
=& \hbar c|\alpha |^{2}\frac{\int_{0}^{\infty }k^{5}l(k)^{2}dk}{%
\int_{0}^{\infty }k^{4}l(k)^{2}dk}\,,
\end{split}
\label{eq:Eavg_tmp}
\end{equation}%
and for a lineshape specified by $l(k)$ (\ref{luse}), we find 
\begin{equation}
\left\langle \mathfrak{E}\right\rangle =\frac{\pi ^{4}}{30\zeta (3)}%
k_{B}T|\alpha |^{2},
\end{equation}
giving $\left\langle \mathfrak{E}\right\rangle =1.34|\alpha |^{2}$ $eV$ at $%
T=5777$ $K$. \ As expected, the energy of the pulse depends on the amplitude
of the coherent state $|\alpha |$. This amplitude can in principle be chosen
arbitrarily, and the mixture will yield an equal-space-point, first-order
correlation function equivalent to that of thermal light provided that the
product $p|\alpha |^{2}$ of $|\alpha |^{2}$ with the weighting density $p$
is given by (\ref{eq:p_th_pulse}). This condition also ensures that such
statistical mixture has the same energy density and photon density as
thermal light. \ The spread in energy of the thermal pulse can be
characterized by $\langle \sigma _{\mathfrak{E}}\rangle =\sqrt{\langle 
\mathfrak{E}^{2}\rangle -\langle \mathfrak{E}\rangle ^{2}}$, which is given by 
\begin{equation}\label{sigmaE}
\begin{split}
\langle \sigma _{\mathfrak{E}}\rangle =& \left( (\hbar c)^{2}|\alpha |^{2}%
\frac{\int_{0}^{\infty }k^{6}l(k)^{2}dk}{\int_{0}^{\infty }k^{4}l(k)^{2}dk}%
\right) ^{\frac{1}{2}} \\
=& \sqrt{\frac{12\zeta (5)}{\zeta (3)}}|\alpha |\,k_{B}T,
\end{split}%
\end{equation}%
with details given in Appendix B; we find $\langle \sigma _{\mathfrak{E}%
}\rangle =1.608\left\vert \alpha \right\vert $ $eV$ at $T=5777$ $K$.

While thermal light has no directionality and is isotropic \cite%
{MandelWolfBook}, each of our thermal pulses has a mean direction of
propagation. \ We can characterize it by identifying the average momentum of
the pulse. \ The momentum operator is 
\begin{equation}
\mathcal{P}=\sum_{\lambda }\int d\mathbf{k\;\hbar k\;}n_{\mathbf{k}\lambda },
\end{equation}%
and the average momentum of a thermal pulse is given by 
\begin{equation}
\begin{split}
\left\langle \mathcal{P}\right\rangle =& \sum_{\lambda }\int d\mathbf{k}%
\langle \alpha f|\hbar \mathbf{k}\,n_{\mathbf{k}\lambda }|\alpha f\rangle
\label{momentum} \\
=& \hbar |\alpha |^{2}\left( \frac{C_{1}+C_{3}}{C_{0}+C_{2}}\right) \frac{%
\int_{0}^{\infty }k^{5}l(k)^{2}dk}{\int_{0}^{\infty }k^{4}l(k)^{2}dk}\hat{%
\mathbf{m}}\,   \\
=& \frac{C_{1}+C_{3}}{C_{0}+C_{2}}\frac{\pi ^{4}}{30\zeta (3)}\frac{k_{B}T}{c%
}|\alpha |^{2}\hat{\mathbf{m}},  
\end{split}%
\end{equation}
where in the last line we have used (\ref{luse}), and details are presented
in Appendix B. \ As expected, the average momentum is along the mean
propagation direction $\hat{\mathbf{m}}$. We can push the evaluation further
by assuming a particular form for $\upsilon (x)$, which should be defined on 
$[-1:1]$ and is peaked at $x=1$; we take 
\begin{equation}
\upsilon (x)=\theta (x)\theta (1-x)\mathrm{e}^{-(x-1)^{2}/\gamma ^{2}},
\label{vuse}
\end{equation}%
where $\theta (x)$ is the Heaviside function, and the smaller the
dimensionless parameter $\gamma $ the more the propagation of momentum in
the pulse is centered about $\mathbf{\hat{m}}$. \ Numerically evaluating the
integrals for $\gamma =0.1$ gives $\left\langle \mathcal{P}\right\rangle
\approx 1.3$ $|\alpha |^{2}\hat{\mathbf{m}}$ $eV/c$. This remains almost constant for $\gamma $ in the
range $\left[ 0.01;0.1\right] $. We have verified that we obtain similar
behaviour with different forms for $\upsilon (x)$. \ 

\begin{figure}[t]
\includegraphics[width=0.8\columnwidth]{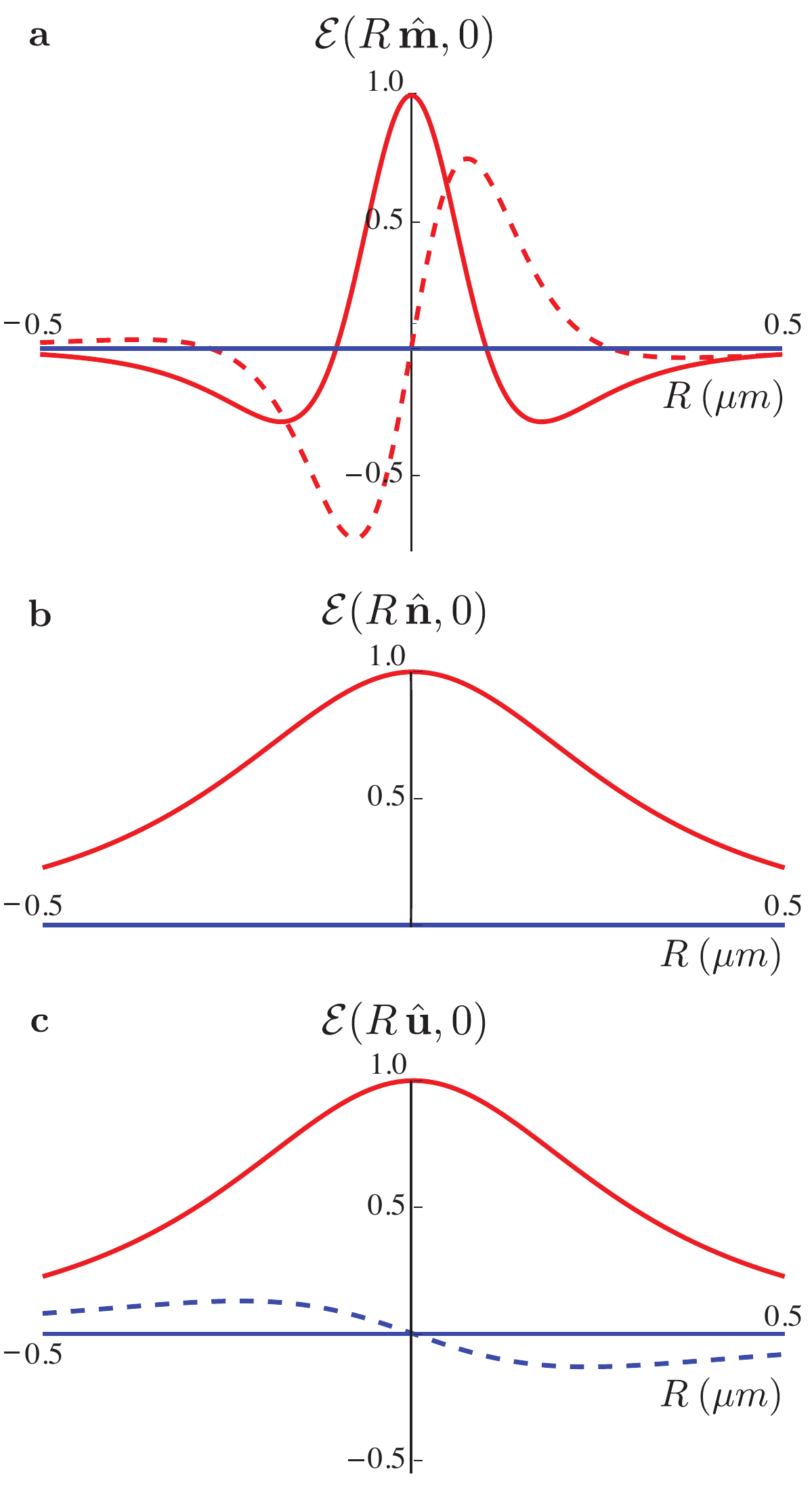} \vspace*{1em}
\caption{Real (dashed lines) and imaginary (plain lines) parts of the
positive-frequency part of the electric field components $\mathcal{E}_{u}(R\,\hat{\mathbf{i}},0) 
$ (red), $\mathcal{E}_{m}(R\,\hat{\mathbf{i}},0) $ (blue) at
position $\mathbf{R}\cdot \hat{\mathbf{i}}$ along the system coordinates $%
\hat{\mathbf{i}}\equiv \hat{\mathbf{m}}$ (a), $\hat{\mathbf{n}}$ (b) and $%
\hat{\mathbf{u}} $ (c). }
\label{fig:E_thermal_pulse}
\end{figure}

The variance in momentum is characterized by the dyadic $\langle \sigma _{%
\mathcal{P}}\rangle ^{2}=\left\langle \mathcal{PP}\right\rangle
-\left\langle \mathcal{P}\right\rangle \left\langle \mathcal{P}\right\rangle 
$. The strategy for evaluating this is sketched in Appendix B; the result is 
\begin{equation}
\begin{split}
\langle \sigma _{\mathcal{P}}\rangle ^{2}& =\hbar ^{2}|\alpha |^{2}\frac{\pi 
}{4}\mathcal{N}^{2}\int_{0}^{\infty }dk \: k^{6}l(k)^{2}\Big(2(C_{4}+C_{2})%
\mathbf{\hat{m}\hat{m}} \\
& +(C_{0}+2C_{2}-3C_{4})\hat{\mathbf{n}}\hat{\mathbf{n}}%
+(3C_{0}-2C_{2}-C_{4})\hat{\mathbf{u}}\hat{\mathbf{u}}\Big).
\end{split}
\label{momvariance}
\end{equation}%
Taking (\ref{vuse}) with $\gamma =0.1$ and $T=5777$ $K$, for example, we
find 
\begin{equation}
\langle \sigma _{\mathcal{P}}\rangle ^{2}\approx 1.19 \left\vert \alpha
\right\vert ^{2}(\uv{m}\uv{m}+0.079 \mathbf{\hat{n}\hat{n}}+0.084%
 \mathbf{\hat{u}\hat{u}})\:(eV/c)^{2}.  \label{special}
\end{equation}%
Yet regardless of the temperature $T$ or the form of $v(x)$, since the
vectors $\mathbf{\hat{m}}$, $\mathbf{\hat{n}}$, and $\mathbf{\hat{u}}$ are
mutually orthogonal we can take the square root of (\ref{momvariance}) by
simply taking the square root of each of the components. \ In the special
case of (\ref{special}) we find 
\begin{equation}
\langle \sigma _{\mathcal{P}}\rangle \approx 1.09 \left\vert
\alpha \right\vert\left( \mathbf{\hat{m}\hat{m%
}}+0.28 \mathbf{\hat{n}\hat{n}}+0.29 \mathbf{\hat{u}\hat{u}}\right)  \;eV/c.
\end{equation}%
The largest spread of momentum is in the direction $\mathbf{\hat{m}}$ the
pulse is nominally propagating; the spreads in the other two directions are
almost but not identically the same. \ 

\begin{figure}[h]
\includegraphics[width=0.8\columnwidth]{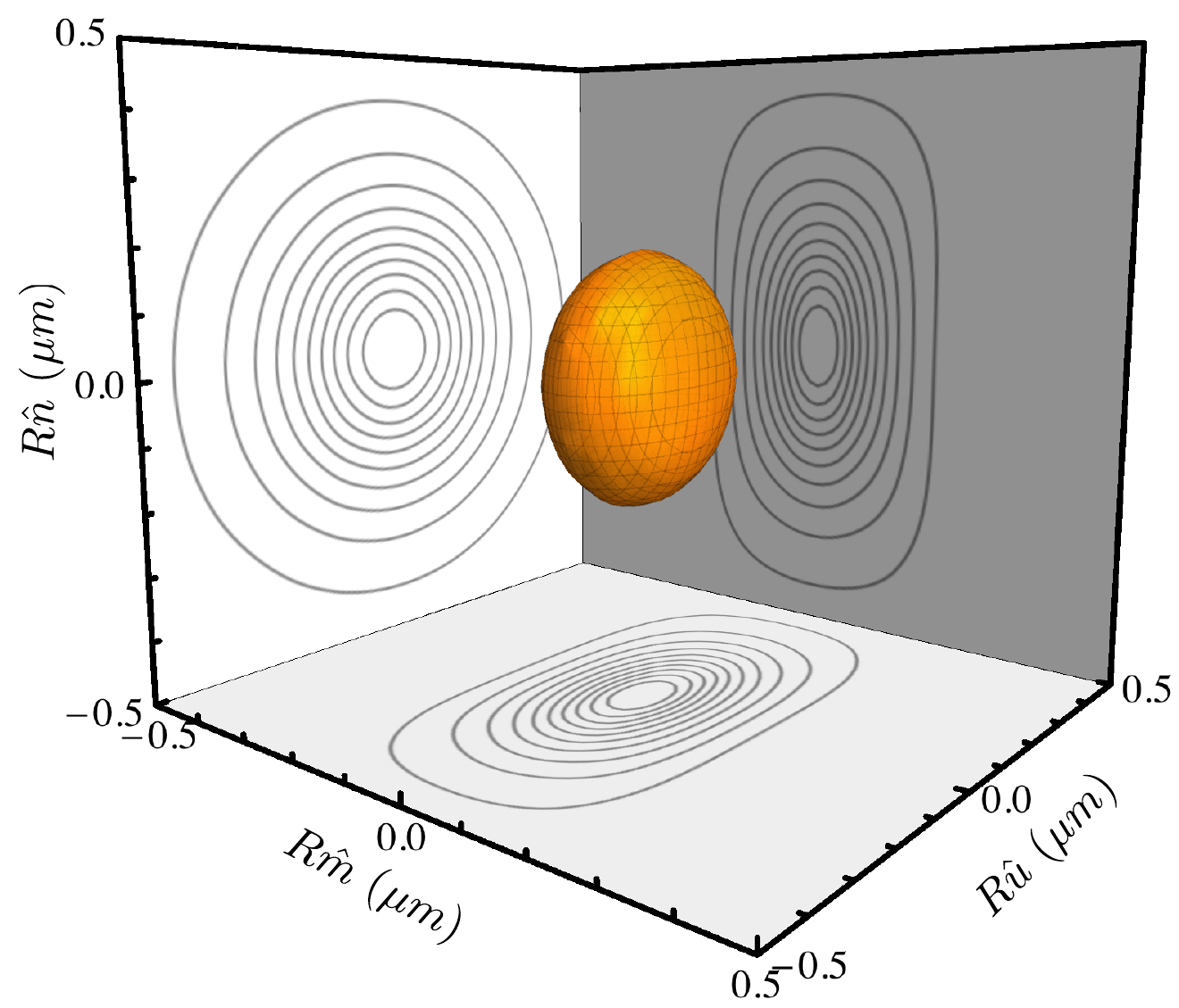}\newline
\caption{Contour plots of the intensity for the `thermal' pulse (Eq. \ref{eq:I_thermal_pulse}). 
The 3D zone delimitates the half maximum region. }
\label{fig:I_thermal_pulse}
\end{figure}

Finally, while the energy density of thermal radiation is uniform, our
pulses are localized in space. \ In Appendix B we give the details for
calculating the expectation value (\ref{E_broadband}) of the
positive-frequency part of the electric field. \ Since this quantity depends
only on $\mathbf{r-r}_{\rm o}\equiv \mathbf{R}$ we relabel the left-hand-side of
(\ref{E_broadband}) as $\mathcal{E}\left( \mathbf{R},t\right) $, where the
pulse is further identified by the direction $\mathbf{\hat{m}}$ of the
expectation value of its momentum and the unit vector $\mathbf{\hat{n}}$. \
Since the electric field in the pulse has no component in the $\mathbf{\hat{n%
}}$ direction, the nonvanishing components will only be those along $\mathbf{%
\hat{m}}$ $\left( \mathcal{E}_{m}\left( \mathbf{R},t\right) \right) $ and
along $\mathbf{\hat{u}=\hat{m}\times \hat{n}}$ $\left( \mathcal{E}_{u}\left( 
\mathbf{R},t\right) \right) $, 
\begin{equation}
\mathcal{E}\left( \mathbf{R},t\right) =\mathcal{E}_{u}\left( \mathbf{R}%
,t\right) \,\hat{\mathbf{u}}+\mathcal{E}_{m}\left( \mathbf{R},t\right) \,%
\hat{\mathbf{m}}.  \label{pulsefield}
\end{equation}%
In Fig. \ref{fig:E_thermal_pulse} we show the dependence of the electric
field components as a function of the distance to the centre\ of the pulse $%
R\equiv |\mathbf{r}-\mathbf{r}_{\mathrm{o}}|$. For the figure, we chose $%
\upsilon (x)$ with a Gaussian shape and $\gamma =0.1$, and put $\alpha =1$.
We have investigated the use of shapes different than (\ref{vuse}) for $%
\upsilon (x)$, e.g. $\upsilon (x)=\theta (x)\theta (1-x)\,(1-(x-1)^{2})$,
and find that the shape of the electric field is not significantly different
for these different $\upsilon (x)$ functions, as long as they are peaked at $%
x=1$. \ From (\ref{pulsefield}) we can easily construct an intensity
function,%
\begin{equation}
I(\mathbf{R},t)=|\mathcal{E}_{u}\left( \mathbf{R},t\right) |^{2}+|\mathcal{E}%
_{m}\left( \mathbf{R},t\right) |^{2}.  \label{eq:I_thermal_pulse}
\end{equation}%
In Fig. \ref{fig:I_thermal_pulse} we illustrate the region of space for
which the intensity is half of its maximum value, and gives the contour
plots in 3D space.

\section{Conclusions}

Much of the intuition of researchers in optical physics deals with how
different states of light are related to each other.  Thermal light is well
represented by a mixture of tensor products of coherent states describing
monochromatic radiation, i.e., continuous waves. \ Yet broadband coherent
states, i.e., pulses, are central to resolving the dynamics of physical
processes. The relation between the two states of light involved here is
especially relevant to the study of biological systems such as
photosynthetic complexes, which under natural conditions are excited by
thermal light, but which are probed in the laboratory with ultra-fast pulses.

In previous work  \cite{Chenu2014b} we showed that thermal light cannot be represented as
an incoherent mixture of pulses, in contradiction to the intuition of at
least some.  Here we showed that the intuition can be maintained if interest
is restricted to first-order, equal-space-point correlation functions
representative of linear light-matter interaction. We presented two families
of pulses, mixtures of which can be used to reproduce the features of the
linear interaction of matter with thermal light: (i) pulses with a Gaussian
lineshape, with surprisingly narrow bandwidths, (ii) and "thermal" pulses,
with the coherence time of thermal light. In each case the density operator
describing the mixture of pulses must be improper, in that its trace is not
unity but rather scales with the volume containing the radiation. \ 

The decompositions presented in this paper may prove to be helpful
conceptual tools, and useful in some calculations where localized pulses are
easier to treat than radiation extending over all space.  We also note that
while thermal light is often used as a proxy for sunlight, thermal light and
sunlight are strikingly different in that sunlight carries momentum while
thermal light does not. \ So the kind of decompositions we present here
might be of use in constructing mixtures to properly represent sunlight. \ 

The conclusion of our earlier work  \cite{Chenu2014b} of course remains:  Nonlinear
light-matter interactions involving thermal light, described by higher-order
correlation functions, cannot in general be described with the aid of
mixtures of single pulses of the type presented in this paper. Yet further
work is required to investigate the existence of regimes where the use of
such a mixture may be at least approximately valid. \  This should help in
the development of conceptual and calculational tools for understanding and
designing nonlinear optical experiments to study excitation by thermal
light.

\begin{acknowledgements}
We are grateful to P. Brumer, B. Sanders, A. Steinberg and H. Wiseman for interesting and fruitful discussions. A.C. acknowledges funding from the DFAIT of Canada and the Swiss National Science Foundation, and G.D. Scholes for hosting her. This work was partly supported by the Natural Sciences and Engineering Research Council of Canada. 
Research at Perimeter Institute is supported by the Government of Canada through Industry Canada and by the Province of Ontario through the Ministry of Research and Innovation.
\end{acknowledgements}

\clearpage

\onecolumngrid
\appendix

\section{First-order correlation function for Gaussian-like pulses}

\label{app:G1_Gaussian} We present here the details of the calculation for
the first-order correlation function of the trace-improper mixture composed
of pulses with Gaussian lineshape (\ref{gaussianG}). \ In addition to the
system of coordinates $\{\hat{\mathbf{e}}_{1\mathbf{k}},\hat{\mathbf{e}}_{2%
\mathbf{k}},\hat{\mathbf{k}}\}$, it is useful to have two sets of three
mutually orthogonal unit vectors available for the derivations here and in
the following appendices, and we illustrate them in Fig. (\ref%
{fig:coordinate}). \ One of our sets is $\left\{ \mathbf{\hat{e}}_{1\mathbf{m%
}},\mathbf{\hat{e}}_{2\mathbf{m}},\mathbf{\hat{m}}\right\} $, where $\mathbf{%
\hat{m}}$ identifies the direction of $\mathbf{k}_{\mathrm{o}}$ and $\mathbf{%
\hat{e}}_{1\mathbf{m}}$ and $\mathbf{\hat{e}}_{2\mathbf{m}}$ are two unit
vectors orthogonal to each other and to $\mathbf{\hat{m}}$, such that $\hat{%
\mathbf{e}}_{1\mathbf{m}}\times \hat{\mathbf{e}}_{2\mathbf{m}}=\hat{\mathbf{m%
}}$. \ The pulses $\mathcal{E}_{i}^{g}(\mathbf{r},t)$, given by (\ref{E_gaussian}), 
also involve the vector $\mathbf{\hat{n}}$, which is
perpendicular to $\mathbf{\hat{m}}$. \ It lies in the plane of $\mathbf{\hat{%
e}}_{1\mathbf{m}}$ and $\mathbf{\hat{e}}_{2\mathbf{m}}$, and we specify it as%
\begin{equation}
\hat{\mathbf{n}}=\hat{\mathbf{e}}_{1\mathbf{m}}\cos \Psi +\hat{\mathbf{e}}_{2%
\mathbf{m}}\sin \Psi .  \label{nhatwrite}
\end{equation}%
Our second set of unit vectors is $\left\{ \mathbf{\hat{n},\hat{u},\hat{m}}%
\right\} $ where $\mathbf{\hat{u}}\equiv $ $\mathbf{\hat{m}\times \hat{n}}$.
\ 

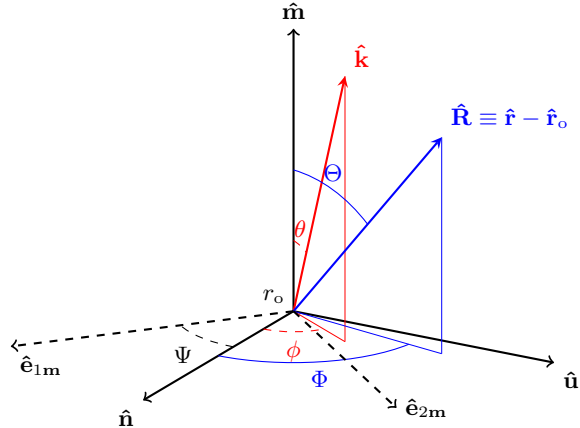
\begin{figure}[b]
\tdplotsetmaincoords{70}{120} \pgfmathsetmacro{\kvec}{1} \pgfmathsetmacro{%
\thetavec}{20} \pgfmathsetmacro{\phivec}{60} \pgfmathsetmacro{\rvec}{1} %
\pgfmathsetmacro{\xivec}{40} \pgfmathsetmacro{\psivec}{80}
\pgfmathsetmacro{\rx}{\rvec *sin(\xivec)* cos(\psivec)} \pgfmathsetmacro{%
\ry}{\rvec *sin(\xivec)* sin(\psivec)} 
\pgfmathsetmacro{\alphavec}{-40} 
\begin{tikzpicture}[scale=4,tdplot_main_coords]
    \coordinate (O) at (0,0,0);
     \draw(0,0,0) -- (0,0,0) node[anchor=south east]{$r_{\rm o}$};
    \draw[thick,->] (0,0,0) -- (1,0,0) node[anchor=north east]{$\uv{n}$};
    \draw[thick,->] (0,0,0) -- (0,1,0) node[anchor=north west]{$\uv{u}$};
    \draw[thick,->] (0,0,0) -- (0,0,1) node[anchor=south]{$\uv{m}$};
    \tdplotsetcoord{K}{\kvec}{\thetavec}{\phivec}
    \draw[-stealth,color=red,thick] (O) -- (K) node[above right] {$\uv{k}$};
    \draw[red, very thin] (O) -- (Kxy);
    \draw[red, very thin] (K) -- (Kxy);
    \tdplotdrawarc[dashed, red]{(O)}{0.2}{0}{\phivec}{anchor=north, red}{$\phi$}
    \tdplotsetthetaplanecoords{\phivec}
    \tdplotdrawarc[tdplot_rotated_coords, dashed, red]{(0,0,0)}{0.25}{0}%
        {\thetavec}{anchor=south}{$\theta$}
    \coordinate (O) at (0,0,0);
    \tdplotsetcoord{R}{\rvec}{\xivec}{\psivec}
    \draw[-stealth,color=blue,thick] (O) -- (R) node[above right] {$\uv{R}\equiv\uv{r}-\uv{r}_{\mathrm{\rm o}}$};
    \draw[blue, very thin] (O) -- (Rxy);
    \draw[blue, very thin] (R) -- (Rxy);
    \tdplotdrawarc[blue]{(O)}{0.5}{0}{\psivec}{anchor=north}{$\Phi$}
    \tdplotsetthetaplanecoords{\psivec}
    \tdplotdrawarc[tdplot_rotated_coords, blue]{(0,0,0)}{0.5}{0}%
        {\xivec}{anchor=south}{$\Theta$}
\tdplotsetrotatedcoords{\alphavec}{0}{0}
\tdplotsetrotatedcoordsorigin{(O)}
\draw[thick,dashed,tdplot_rotated_coords,->] (0,0,0) -- (1,0,0) node[anchor=north west]{$\uv{e}_{1 \bm}$};
\draw[thick,dashed,tdplot_rotated_coords,->] (0,0,0) -- (0,1,0) node[anchor=west]{$\uv{e}_{2 \bm}$};
    \tdplotdrawarc[dashed]{(O)}{0.4}{0}{\alphavec}{anchor=north east}{$\Psi$}
\end{tikzpicture}
\caption{Representation of selected vectors and angles used in the
derivations. The two sets of mutually orthogonal unit vectors $\{ \hat{%
\mathbf{e}}_{1 \mathbf{m}},\hat{\mathbf{e}}_{2 \mathbf{m}},\hat{\mathbf{m}}%
\} $ and $\{\hat{\mathbf{n}},\hat{\mathbf{u}},\hat{\mathbf{m}}\}$ each form
a direct orthonormal system. The centre is at position $r_{\mathrm{o}}$. }
\label{fig:coordinate}
\end{figure}

We are now ready to begin. \ Once the expression for $\mathcal{E}_{i}^{%
\mathrm{g}}(\mathbf{r},t)$ is used in (\ref{gaussianG}), the integration
over the positions $\mathbf{r}_{\mathrm{o}}$ of the pulses can be done
immediately using $\int \mathrm{e}^{i(%
\mathbf{k}^{\prime }-\mathbf{k})\cdot (\mathbf{r}-\mathbf{r}_{\mathrm{o}})}d%
\mathbf{r}_{\mathrm{o}}=\left( 2\pi \right) ^{3}\delta (\mathbf{k}-\mathbf{k}%
^{\prime })$, we find 
\begin{equation}  \label{eq:G1_first}
\begin{split}
G_{ij}^{(1){g}}(\mathbf{r}t;\mathbf{r}t^{\prime }) =(2\pi )^{3}|i\alpha 
\mathcal{N}^{g}|^{2}\int \,d\mathbf{k}_{\mathrm{o}}\,d\mathbf{k}\,p(k_{%
\mathrm{o}})\,\frac{\hbar \,c\,k}{16\pi ^{3}\epsilon _{0}}\mathrm{e}^{-\frac{%
|\mathbf{k}-\mathbf{k}_{\mathrm{o}}|^{2}}{\sigma ^{2}}}\mathrm{e}^{-{i}%
\omega _{k}(t^{\prime }-t)}T_{ij}(\mathbf{k,\hat{m}}),
\end{split}%
\end{equation}%
where 
\begin{eqnarray}
T_{ij}(\mathbf{k,\hat{m}}) &=&\int_{0}^{2\pi }d\Psi \left( \hat{\mathbf{i}}%
\cdot (\mathbf{k}\times \hat{\mathbf{n}})\right) \,\left( \hat{\mathbf{j}}%
\cdot (\mathbf{k}\times \hat{\mathbf{n}})\right)  \label{workT} \\
&=&\varepsilon _{i\eta \mu }\varepsilon _{j\sigma \nu }k_{\eta }k_{\sigma
}\int_{0}^{2\pi }d\Psi \;n_{\mu }n_{\nu }  \notag \\
&=&\pi \varepsilon _{i\eta \mu }\varepsilon _{j\sigma \nu }k_{\eta
}k_{\sigma }\left( \delta _{\mu \nu }-m_{\mu }m_{\nu }\right)  \notag \\
&=&\pi \left( \delta _{ij}k^{2}-k_{i}k_{j}\right) -\pi \varepsilon _{i\eta
\mu }\varepsilon _{j\sigma \nu }k_{\eta }k_{\sigma }m_{\mu }m_{\nu }\, . 
\notag
\end{eqnarray}%
In the first line of (\ref{workT}) we have written the integral $d\mathbf{%
\hat{n}}$ appearing in (\ref{gaussianG}) as an integral $d\Psi $, with $\Psi 
$ varying from $0$ to $2\pi $; in the second line we have introduced the
Levi-Civita tensor ($\varepsilon _{\alpha \beta \gamma }=1$ if $\{\alpha
,\beta ,\gamma \}$ is a cyclic permutation, $-1$ if the permutation is
anti-cyclic, and 0 if any two indices are equal) to write the cross
products; in the third line we have used (\ref{nhatwrite}) to evaluate 
\begin{eqnarray}
\int_{0}^{2\pi }d\Psi \;\mathbf{\hat{n}\hat{n}} &=&\pi (\hat{\mathbf{e}}_{1%
\mathbf{m}}\hat{\mathbf{e}}_{1\mathbf{m}}+\hat{\mathbf{e}}_{2\mathbf{m}}\hat{%
\mathbf{e}}_{2\mathbf{m}}) \\
&=&\pi (\mathbf{U}-\mathbf{\hat{m}\hat{m})}, \notag
\end{eqnarray}%
where we have used $\mathbf{U}$ to indicate the unit dyadic. \ Finally, in
the last line of (\ref{workT}) we have used the identity $\varepsilon
_{i\eta \mu }\varepsilon _{j\sigma \mu }=\delta _{ij}\delta _{\eta \sigma
}-\delta _{i\sigma }\delta _{j\eta }$ to find 
\begin{equation}
\varepsilon _{i\eta \mu }\varepsilon _{j\sigma \nu }k_{\eta }k_{\sigma
}\delta _{\mu \nu }=\delta _{ij}k^{2}-k_{i}k_{j} .  \label{LCresult}
\end{equation}

With the dependence of $T_{ij}(\mathbf{k,\hat{m})}$ on $\mathbf{k}$ and $%
\mathbf{\hat{m}}$ explicit in (\ref{workT}), we can now integrate over $%
\mathbf{\hat{m}}$, the direction of $\mathbf{k}_{\mathrm{o}}$, keeping $%
\mathbf{k}$ fixed. \ Writing out $|\mathbf{k}-\mathbf{k}_{\mathrm{o}}|^{2}=(%
\mathbf{k}-\mathbf{k}_{\mathrm{o}})\cdot (\mathbf{k}-\mathbf{k}_{\mathrm{o}%
})=k^{2}+k_{\mathrm{o}}^{2}-2kk_{\mathrm{o}}\cos \theta $, where $\theta $
is the angle between $\mathbf{k}$ and $\mathbf{k}_{\mathrm{o}}$, we have
\begin{equation}
\begin{split}
G_{ij}^{(1){g}}(\mathbf{r}t;\mathbf{r}t^{\prime }) =&\frac{(2\pi
)^{3}|\alpha \mathcal{N}^{g}|^{2}\hbar \,c\,}{16\pi ^{3}\epsilon _{0}}\int
\,d\mathbf{k}\,\int k_{\mathrm{o}}^{2}dk_{\mathrm{o}}\;p(k_{\mathrm{o}})\,k\,%
\mathrm{e}^{-(k^{2}+k_{\mathrm{o}}^{2})/\sigma ^{2}}\mathrm{e}^{-{i}%
ck(t^{\prime }-t)}  \label{Gwork1} \\
&\times \int d\mathbf{\hat{m}\;}T_{ij}(\mathbf{k,\hat{m}})\mathrm{e}^{2kk_{%
\mathrm{o}}\cos \theta /\sigma ^{2}}.  
\end{split}
\end{equation}%
Using the unit vectors $\mathbf{\hat{e}}_{1\mathbf{k}}$ and $\mathbf{\hat{e}}%
_{2\mathbf{k}}$ appearing in (\ref{helicity}) and defined below that
equation, we can introduce an angle $\bar{\phi}$ between $\mathbf{\hat{e}}_{1%
\mathbf{k}}$ and the projection of $\mathbf{\hat{m}}$ on the plane defined
by $\mathbf{\hat{e}}_{1\mathbf{k}}$ and $\mathbf{\hat{e}}_{2\mathbf{k}}$,
such that in the usual way $\mathbf{\hat{m}}=\mathbf{\hat{k}}\cos \theta +%
\mathbf{\hat{e}}_{1\mathbf{k}}\sin \theta \cos \bar{\phi}+\mathbf{\hat{e}}_{2%
\mathbf{k}}\sin \theta \sin \bar{\phi}$ and $d\mathbf{\hat{m}=}\sin \theta
d\theta d\bar{\phi}$. \ (Note that neither $\bar{\phi}$, nor the vectors $%
\mathbf{\hat{e}}_{1\mathbf{k}}$ and $\mathbf{\hat{e}}_{2\mathbf{k}}$, are
shown in Fig. \ref{fig:coordinate}). \ To find the integral over $\bar{\phi}$
of $T_{ij}(\mathbf{k,\hat{m}})$ we need \ 
\begin{equation}
\int_{0}^{2\pi }d\bar{\phi}\;\mathbf{\hat{m}\hat{m}}=\pi \mathbf{U}\sin
^{2}\theta +\pi \mathbf{\hat{k}\hat{k}}(2\cos ^{2}\theta -\sin ^{2}\theta ),
\label{phibarwork}
\end{equation}%
which follows from a straight-forward integration, and using $\mathbf{U=\hat{%
k}\hat{k}+\hat{e}}_{1\mathbf{k}}\mathbf{\hat{e}}_{1\mathbf{k}}+\mathbf{\hat{e%
}}_{2\mathbf{k}}\mathbf{\hat{e}}_{2\mathbf{k}}$. \ Using (\ref{workT},\ref%
{phibarwork}) we can then find 
\begin{equation}
\int_{0}^{2\pi }d\bar{\phi}\;T_{ij}(\mathbf{k,\hat{m}})=2\pi ^{2}(\delta
_{ij}k^{2}-k_{i}k_{j})\left(1-\frac{1}{2}\sin ^{2}\theta \right),  \label{phibarresult}
\end{equation}%
where the second term in (\ref{phibarwork}) makes no contribution when (\ref%
{workT}) is used in (\ref{phibarresult}), since $\mathbf{k\times k}=0$.
Using (\ref{phibarresult}) in the last line of (\ref{Gwork1}) we see that
only the remaining integral involving $\sin \theta d\theta $ needs to be
done in that line. Defining $a=2kk_{\mathrm{o}}/\sigma ^{2}$, we can finish
that evaluation by noting that \ \ \ 
\begin{subequations}
\begin{align}
\int_{0}^{\pi }\sin \theta \,\mathrm{e}^{a\cos \theta }d\theta =&
\int_{-1}^{1}\mathrm{e}^{ax}dx=2\frac{\sinh (a)}{a} \\
\int_{0}^{\pi }\sin ^{3}\theta \,\mathrm{e}^{a\cos \theta }d\theta =&
\int_{-1}^{1}(1-x^{2})\mathrm{e}^{ax}dx=4\frac{a\cosh (a)-\sinh (a)}{a^{3}}.
\end{align}%
Using the expression (\ref{eq:n_Gaussian}) for the normalization constant $%
\mathcal{N}^{g}$ we then find 
\end{subequations}
\begin{equation}
\begin{split}
G_{ij}^{(1){g}}(\mathbf{r}t;\mathbf{r(}t+\tau ))=& \frac{(2\pi )^{3}\hbar
c|\alpha |^{2}}{16\pi ^{3}\epsilon _{0}\sqrt{\pi }\sigma ^{3}}\int d\mathbf{k%
}\int_{0}^{\infty }k_{\mathrm{o}}^{2}dk_{\mathrm{o}}p(k_{\mathrm{o}})\frac{k%
}{(\sigma ^{2}+k_{\mathrm{o}}^{2})}\,\mathrm{e}^{-\frac{k^{2}+k_{\mathrm{o}%
}^{2}}{\sigma ^{2}}}\mathrm{e}^{-{i}ck\tau } \\
&\times \Big[\pi \left( \delta _{ij}k^{2}-k_{i}k_{j}\right) \left( 4\frac{%
\sinh a}{a}-4\frac{a\cosh a-\sinh a}{a^{3}}\right) \Big].
\end{split}%
\end{equation}%
We can further integrate over the orientation $\Omega $ of $d\mathbf{k}%
=k^{2}dkd\Omega $ using $\int \left( \delta _{ij}k^{2}-k_{i}k_{j}\right)
d\Omega =\frac{8\pi }{3}\delta _{ij}k^{2}$; writing out and combining the
terms the correlation function for the mixture of Gaussian pulses becomes (%
\ref{Gaussianresult}). 

\section{Characterization of the pulses with a more general line shape
(thermal pulses)}

\label{app:broad}

\subsection{First-order correlation function}

\label{app:G1_pulse} We now turn to the evaluation of the first-order
correlation function for the mixture (\ref{eq:g1_b-def}), where $\mathcal{E}%
_{i}^{b}(\mathbf{r},t)$ denotes the $i$th component of the classical electric field given by (\ref%
{eq:E_general}) with $L\rightarrow L^{b}$ (\ref{f_broadband}) and the
normalization factor $\mathcal{N}\rightarrow \mathcal{N}^{\mathrm{b}}$ (\ref%
{n_broadband}). \ As in the example of Gaussian pulses considered above,
once the expression for $\mathcal{E}_{i}^{b}(\mathbf{r},t)$ is used in (\ref%
{eq:g1_b-def}), the integration over the positions $\mathbf{r}_{\mathrm{o}}$
of the pulses can be done immediately 
and we find 
\begin{eqnarray}  \label{eq:gaussian_int_0}
G_{ij}^{(1){b}}(\mathbf{r}t;\mathbf{r}(t+\tau )) ={}(2\pi )^{3}|\alpha 
\mathcal{N}^{b}|^{2}\int d\mathbf{k}\:p\,\left( \frac{\hbar ck}{16\pi
^{3}\epsilon _{0}}\right) \,l(k)^{2}\mathrm{e}^{-ick\tau }\int d\hat{\mathbf{%
m}}\,(\upsilon (\hat{\mathbf{k}}\cdot \hat{\mathbf{m}}))^{2}\,T_{ij}(\mathbf{%
k,\hat{m}})\,.
\end{eqnarray}%
Again defining $\theta $ and $\bar{\phi}$ as in the analysis of Gaussian
pulses, we have $\upsilon (\hat{\mathbf{k}}\cdot \hat{\mathbf{m}})=\upsilon
(\cos \theta )$, independent of $\bar{\phi}$, and so%
\begin{eqnarray}  \label{bTintegral}
\int d\hat{\mathbf{m}}\,(\upsilon (\hat{\mathbf{k}}\cdot \hat{\mathbf{m}}%
))^{2}\,T_{ij}(\mathbf{k,\hat{m}}) &=&\int_{0}^{\pi }\left[ \int_{0}^{2\pi
}T_{ij}(\mathbf{k,\hat{m}})d\bar{\phi}\right] \upsilon (\cos \theta )\sin
\theta d\theta \\
&=&\pi ^{2}(\delta _{ij}k^{2}-k_{i}k_{j})(C_{0}+C_{2}),  \notag
\end{eqnarray}%
where we have used (\ref{phibarresult}) and the definitions of $C_n$ (\ref%
{Cndef}). \ Using (\ref{bTintegral}) in (\ref{eq:gaussian_int_0}), we can
integrate over the direction $\mathbf{\hat{k}}$ by writing $d\mathbf{k}%
=k^{2}dkd\Omega $, and recalling that $\int k_{i}k_{j}d\Omega =\frac{4\pi }{3%
}k^{2}\delta _{ij}$. \ Finally, using the result (\ref{n_broadband}) for $%
\mathcal{N}^{\mathrm{b}}$, we find (\ref{broadbandresult}). \

\subsection{Electric field}

\label{app:E_pulse} In the notation defined in the text before (\ref%
{pulsefield}), we have
\begin{equation} \label{eq:E_thermal-pulse_general}
\mathcal{E(}\mathbf{R},t)=i\mathcal{N}^{b}\alpha \int d\mathbf{k}\;\sqrt{%
\frac{\hbar \omega _{k}}{16\pi ^{3}\epsilon _{0}}}(\mathbf{k\times \hat{n}}%
)\,l(k)\,\upsilon (\hat{\mathbf{k}}\cdot \hat{\mathbf{m}})e^{i\mathbf{k\cdot R}%
}e^{-i\omega _{k}t}.
\end{equation}%
Putting $\mathbf{R}=R(\sin \Theta \cos \Phi \,\hat{\mathbf{n}}+\sin \Theta
\sin \Phi \,\hat{\mathbf{u}}+\cos \Theta \,\hat{\mathbf{m}})$ (\textit{cf.}
Fig. \ref{fig:coordinate}), we hold $\Theta \,$\ and $\Phi $ fixed and
perform the integrations over the angles $\theta $ and $\phi $ that define
the direction of the wave vector $\mathbf{k}$ relative to the nominal
direction of propagation $\hat{\mathbf{m}}$. The integration over $\phi $
can be evaluated analytically: 
\begin{equation}\label{phiintegral}
\int_{0}^{2\pi }d\phi \,\mathrm{e}^{-i\mathbf{k}\cdot \mathbf{R}}(\mathbf{k%
}\times \hat{\mathbf{n}})=2\pi k\,\mathrm{e}^{-ikR(\cos \theta \cos \Theta )}\left[ J_{0}(kR\sin
\theta \sin \Theta )\cos \theta \,\hat{\mathbf{u}}+i\sin \theta \sin \Phi
J_{1}(kR\sin \theta \sin \Theta )\,\hat{\mathbf{m}}\right] \,,
\end{equation}
where we have written out the dot and cross products appearing in the
integrand, and recognized integral expressions for the Bessel functions $%
J_{0}$ and $J_{1}$ \cite{Abramowitz1965a}. \ Using (\ref{phiintegral}) in (\ref%
{eq:E_thermal-pulse_general}) and recalling that for a given pulse both $%
\mathbf{\hat{m}}$ and $\mathbf{\hat{u}}$ are fixed, we find (\ref{pulsefield}%
) with 
\begin{eqnarray}\label{eq:pulse_E}
\mathcal{E}_{u}(\mathbf{R},t)=& {}2\pi i\alpha \mathcal{N}^{\mathrm{b}%
}\int_{0}^{\infty }k^{2}dk\sqrt{\frac{\hbar ck}{16\pi ^{2}\epsilon _{0}}}%
\,k\,l(k)\,\mathrm{e}^{-ickt}\int_{-1}^{1}dx\,\upsilon (x)\,\mathrm{e}%
^{-ikR(x\cos \Theta )}J_{0}(kR\sqrt{1-x^{2}}\sin \Theta )x
 \\
\mathcal{E}_{m}(\mathbf{R},t)=& -2\pi \alpha \mathcal{N}^{\mathrm{b}%
}\int_{0}^{\infty }k^{2}dk\sqrt{\frac{\hbar ck}{16\pi ^{2}\epsilon _{0}}}%
\,k\,l(k)\,\mathrm{e}^{-ickt}\int_{-1}^{1}dx\,\upsilon (x)\,\mathrm{e}%
^{-ikR(x\cos \Theta )}\sqrt{1-x^{2}}\sin \Phi J_{1}(kR\sqrt{1-x^{2}}\sin
\Theta ) \notag
\end{eqnarray}
where we have changed the variable such as $x=\cos \theta $.

Defining a particular shape for $\upsilon(x)$ (\ref{vuse}) allows us to
numerically evaluate the integrals, as presented in Fig. \ref%
{fig:E_thermal_pulse}.

\subsection{Wave packet averages}

As in the text, for the sake of readability we omit the superscript $b$ and
the subscripts $\{\mathbf{r}_{\mathrm{o}}s\}$ in the following, keeping in
mind that $\alpha $, $f$ and $\mathcal{N}$ (\ref{n_broadband}), 
\begin{equation}
\mathcal{N}=\left[ \pi (C_{0}+C_{2})\int_{0}^{\infty }k^{4}l(k)^{2}dk\right]
^{-\frac{1}{2}}  \label{Nuse}
\end{equation}%
are given for the thermal pulses in particular, with the spectral components
of $f$ (\ref{spectralbroadband},\ref{f_broadband}) labeled as $f_{\mathbf{k}%
\lambda }$,%
\begin{equation}
f_{\mathbf{k\lambda }}=\mathcal{N\;}l(k)\,\upsilon (\hat{\mathbf{k}}\cdot 
\hat{\mathbf{m}})\,(\mathbf{e}_{\mathbf{k\lambda }}^{\ast })\cdot (\mathbf{%
k\times \hat{n}})\,\mathrm{e}^{-i\mathbf{k\cdot r}_{\mathrm{o}}}
\label{spectraluse}
\end{equation}%
and the dependence of $f_{\mathbf{k}\lambda }$ on the pulse parameters is
kept implicit.

\subsubsection{Average momentum}

The average momentum for the wave packet defined by Eq. (\ref{eq:ket_pulse})
is: 
\begin{eqnarray}
\left\langle \mathcal{P}\right\rangle &=& \sum_{\lambda }\int d\mathbf{k}%
\langle \alpha f|_{{}}\hbar \mathbf{k}\,{n}_{\mathbf{k}\lambda }|\alpha
f\rangle _{{}} \\
&=& \hbar |\alpha |^{2}\int d\mathbf{k}\,\mathbf{k}\sum_{\lambda }|f_{\mathbf{%
k}\lambda }|^{2} \notag \\
&=& \hbar |\alpha |^{2}\int d\mathbf{k}|\mathcal{N}l(k)|^{2}\mathbf{k}(%
\mathbf{k}\times \hat{\mathbf{n}})\cdot (\mathbf{k}\times \hat{\mathbf{n}}%
)|\upsilon (\hat{\mathbf{k}}\cdot \hat{\mathbf{m}})|^{2}, \notag
\label{eq:avg_mom_general}
\end{eqnarray}%
where in moving from the second to the third line we have used (\ref{kcrossn}%
). \ Using the coordinate system from Fig. (\ref{fig:coordinate}), we can
evaluate the cross products and the angular integration: 
\begin{equation}
\int_{0}^{2\pi }d\phi \:\mathbf{k}(\mathbf{k}\times \hat{\mathbf{n}})\cdot (%
\mathbf{k}\times \hat{\mathbf{n}})=\pi k^{3}\cos \theta (2\cos ^{2}\theta
+\sin ^{2}\theta )\hat{\mathbf{m}},
\end{equation}%
and so 
\begin{equation}
\left\langle \mathcal{P}\right\rangle =\pi \hbar |\alpha
|^{2}\int_{0}^{\infty }dk\,k^{5}|\mathcal{N}l(k)|^{2}(C_{1}+C_{3})\hat{%
\mathbf{m}}.
\end{equation}%
The components of the pulse are characterized by a main direction $\mathbf{%
\hat{m}}$, but the spread of the wave vectors from this direction is
characterized by the function $\upsilon (x)$, and for the form (\ref{vuse})
we can find an analytical solution for the mean momentum: 
\begin{equation}
\left\langle \mathcal{P}\right\rangle =\frac{\pi ^{4}\mathrm{e}^{-\frac{8}{%
\gamma ^{2}}}\left( \mathrm{e}^{\frac{8}{\gamma ^{2}}}\left( \sqrt{2\pi }%
\left( 3\gamma ^{2}+8\right) \text{erf}\left( \frac{2\sqrt{2}}{\gamma }%
\right) -2\gamma \left( \gamma ^{2}+8\right) \right) +2\gamma \left( \gamma
^{2}+4\right) \right) }{30\beta c\zeta (3)\left( \sqrt{2\pi }\left( \gamma
^{2}+8\right) \text{erf}\left( \frac{2\sqrt{2}}{\gamma }\right) -8\gamma
\right) }|\alpha |^{2}\hat{\mathbf{m}}.
\end{equation}%
Numerically evaluating this expression gives the results quoted in the text.

\subsubsection{Standard deviation in energy}

The second central moment on energy is defined as 
\begin{equation}
\langle \sigma_{\mathfrak{E}} \rangle = \sqrt{\langle \mathfrak{E}^2\rangle -
\langle \mathfrak{E}\rangle^2 }.
\end{equation}

We first evaluate the expectation value of the square of the energy, 
\begin{eqnarray}  \label{Esquared}
\langle \mathfrak{E}^{2}\rangle &=& \mathrm{Tr}\left[ \left( \sum_{\lambda
}\int d\mathbf{k}\hbar \omega _{\mathbf{k}}{n}_{\mathbf{k}\lambda }\right)
^{2}|\alpha f\rangle _{{}}\langle \alpha f|_{{}}\right] \\
&=& (\hbar c)^{2}\sum_{\lambda }\int d\mathbf{k}\,k^{2}|\alpha f_{\mathbf{k}%
\lambda }|^{2}+\langle \mathfrak{E}\rangle ^{2}\,, \notag
\end{eqnarray}%
where we have used ${a}_{\mathbf{k}\lambda }{a}_{\mathbf{k}^{\prime }\lambda
^{\prime }}^{\dagger }=\delta _{\lambda \lambda ^{\prime }}\delta (\mathbf{k}%
-\mathbf{k}^{\prime })+{a}_{\mathbf{k}^{\prime }\lambda ^{\prime }}^{\dagger
}{a}_{\mathbf{k}\lambda }$. The variance then is: 
\begin{equation}
\langle \sigma _{\mathfrak{E}}\rangle =\left( (\hbar c)^{2}\sum_{\lambda
}\int d\mathbf{k}\,k^{2}|\alpha f_{\mathbf{k}\lambda }|^{2}\right) ^{\frac{1%
}{2}}\,.
\end{equation}%
We can first sum of the polarization: 
\begin{equation*}
\sum_{\lambda }|f_{\mathbf{k}\lambda }|^{2}=|\mathcal{N}l(k)|^{2}k^{2}(\cos
^{2}\theta +\sin ^{2}\theta \sin ^{2}\phi )|\upsilon (\cos \theta )|^{2},
\end{equation*}%
where we have used (\ref{polsum}), and $\theta $ and $\phi $ are the angles
in Fig. (\ref{fig:coordinate}). Thus the variance becomes 
\begin{eqnarray}
\langle \sigma _{\mathcal{E}}\rangle ^{2}&=& \pi |\alpha \mathcal{N}%
|^{2}\int_{0}^{\infty }k^{2}dk\,(\hbar ck)^{2}|l(k)|^{2}\,k^{2}\int
dx(2x^{2}+(1-x^{2}))|\upsilon (x)|^{2} \\
&=& \pi (\hbar c)^{2}|\alpha \mathcal{N}|^{2}\,(C_{0}+C_{2})%
\int_{-1}^{1}k^{6}|l(k)|^{2}dk, \notag
\end{eqnarray}%
and using (\ref{Nuse}) we find (\ref{sigmaE}).

\subsubsection{Variance and standard deviation on momentum}

The second central moment on momentum is defined as: 
\begin{equation}
\langle \sigma _{\mathcal{P}}\rangle ^{2}=\langle \mathcal{P}^{2}\rangle
-\langle \mathcal{P}\rangle ^{2}.
\end{equation}%
Following the strategy adopted in determining (\ref{Esquared}) we find 
\begin{equation}
\langle \sigma _{\mathcal{P}}\rangle ^{2}=\hbar ^{2}|\alpha
|^{2}\sum_{\lambda }\int d\mathbf{k}\,\mathbf{k}\,\mathbf{k}\,|f_{\mathbf{k}%
\lambda }|^{2}\,,  \label{eq:app_sigmaP}
\end{equation}%
Using (\ref{spectralbroadband}) and expressing $\mathbf{kk}$ in terms of the 
$\theta $ and $\phi $ of Fig. \ref{fig:coordinate}, the integral over $%
\mathbf{k}$ then yields (\ref{momvariance}).

\end{document}